\documentclass[conference]{IEEEtran}
\IEEEoverridecommandlockouts
\usepackage{cite}
\usepackage{amsmath,amssymb,amsfonts}
\usepackage{graphicx}
\usepackage{textcomp}
\usepackage{xcolor}
\usepackage{bm}
\usepackage{url}
\usepackage{multirow}
\usepackage{hyperref}
\usepackage{diagbox}
\usepackage{makecell}
\usepackage{algorithm}
\usepackage{algpseudocode}
\usepackage{boldline}
\usepackage[flushleft]{threeparttable}
\usepackage{pifont}

\begin{document}

\title{Transfer-Based Strategies for \\ Multi-Target Quantum Optimization}

\author{
\IEEEauthorblockN{Vu Tuan Hai\textsuperscript{1},  Bui Cao Doanh\textsuperscript{1}, Le Vu Trung Duong\textsuperscript{1}, Pham Hoai Luan\textsuperscript{1}, and Yasuhiko Nakashima\textsuperscript{1}}
\IEEEauthorblockA{
    \textsuperscript{1} Nara Institute of Science and Technology, 8916–5 Takayama-cho, Ikoma, Nara 630-0192, Japan.\\
Email: vu.tuan\_hai.vr7@naist.ac.jp} 
}

\maketitle

\begin{abstract}
We address the challenge of multi-target quantum optimization, where the objective is to simultaneously optimize multiple cost functions defined over the same quantum search space. To accelerate optimization and reduce quantum resource usage, we investigate a range of strategies that enable knowledge transfer between related tasks. Specifically, we introduce a two-stage framework consisting of a training phase where solutions are progressively shared across tasks and an inference phase, where unoptimized targets are initialized based on prior optimized ones. We propose and evaluate several methods, including warm-start initialization, parameter estimation via first-order Taylor expansion, hierarchical clustering with D-level trees, and deep learning–based transfer. Our experimental results, using parameterized quantum circuits implemented with PennyLane, demonstrate that transfer techniques significantly reduce the number of required iterations while maintaining an acceptable cost value. These findings highlight the promise of multi-target generalization in quantum optimization pipelines and provide a foundation for scalable multi-target quantum optimization.

\end{abstract}


\section{Introduction}

\begin{figure*}
\includegraphics[width=0.99\textwidth]{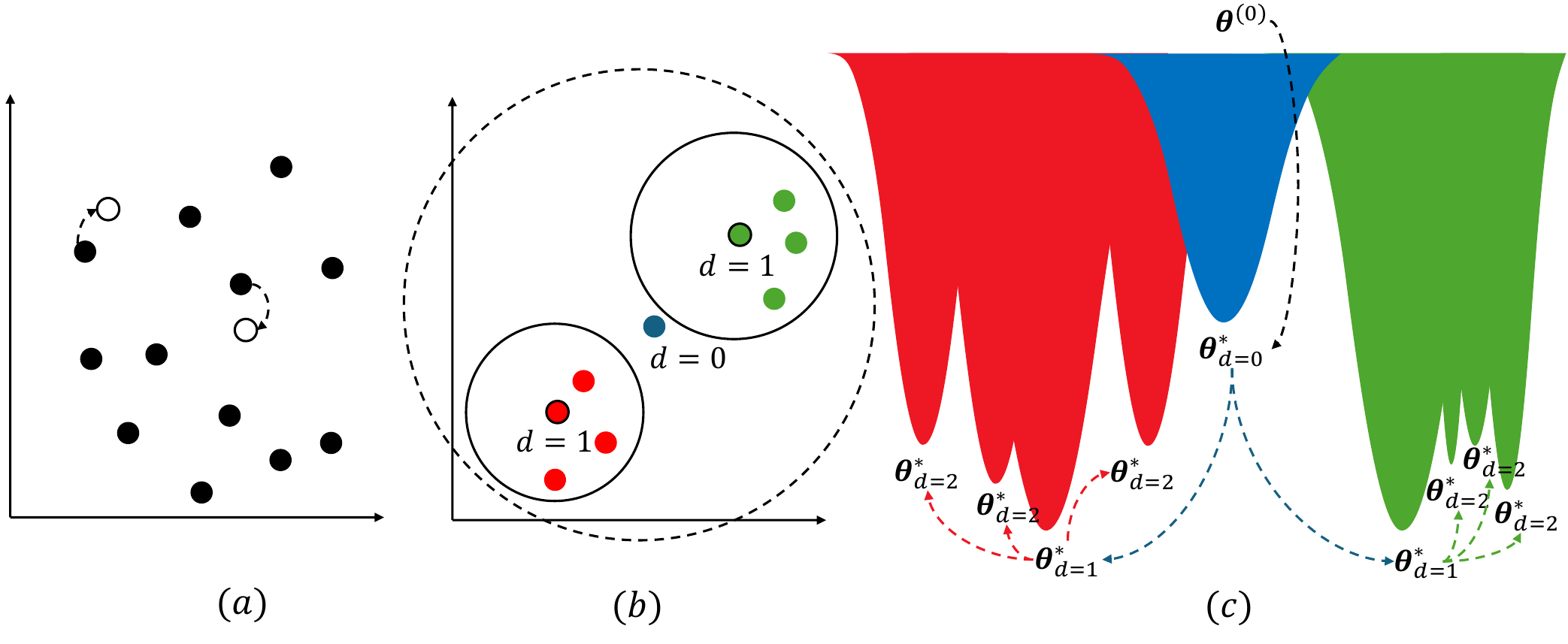}
\caption{Illustration of multi-target optimization space: (a)  Inference stage where there are many existing optimized targets, (b) Training stage where there are no optimized targets, $K$ unoptimized targets as $D$-level tree, and (c) Corresponding landscape for $D$-level tree.}
\label{fig:overview}
\end{figure*}

Quantum computers are expected to solve complex computational problems significantly faster than their classical counterparts, leveraging the properties of superposition and entanglement. This advantage is demonstrated in many fields, from combinatorial optimization~\cite{PhysRevApplied.19.024027, PhysRevLett.129.250502}, cryptography~\cite{doi:10.1137/S0036144598347011} to physical simulations~\cite{Feniou2023}. Among these, quantum optimization has emerged as one of the most practical near-term applications. Quantum optimization promises to accelerate complex targets commonly encountered in classical optimization, such as processing large-scale input~\cite{PhysRevLett.129.250502}. For instance, Quantum Approximate Optimization Algorithms (QAOA) are actively being developed to address combinatorial optimization problems~\cite{PhysRevX.10.021067, PhysRevApplied.19.024027}, while the Variational Quantum Eigensolver (VQE) is widely used for approximating eigenvalues in quantum chemistry~\cite{PhysRevLett.122.230401, Anselmetti_2021}. Moreover, Quantum Neural Networks (QNNs) are increasingly employed to tackle optimization targets within machine learning frameworks~\cite{doi:10.1080/00107514.2014.964942, schuld2021machine, Biamonte2017}. 

The multi-target quantum optimization (MTQO) problem is proposed in \cite{hai2024multi}, which aims to optimize multiple objectives simultaneously, rather than a single objective as in previous works, which aligns with the classical formulation of multi-objective (multi-target) optimization~\cite{NIPS2013_f33ba15e}. Consider a set of $K$ optimization targets denoted as $T_1, T_2, \dots, T_K$, all defined over the same search space  $\mathcal{S}$. Each target $T_k$ is associated with its own cost function $\mathcal{C}(\bm\theta_{(k)}): \mathbb{R}^m \to [0,1]$ where $\bm\theta_{(k)}$ is the $m$-dimensional parameter vector. A target $T_k$ that has the corresponding optimal parameter ($\bm\theta^{*}_{(k)}$) is marked as ``optimized''. The goal of multi-target optimization is to find a set of solutions:

\begin{align}
    \{ \bm\theta^{*}_{(1)}, \bm\theta^{*}_{(2)}, \dots, \bm\theta^{*}_{(K)} \},
\end{align}

with $\bm\theta^{*}_{(k)}$ is the optimal solution for target $T_k$, with a given trainable unitary operator $U(\bm\theta)$. There are two stages of multi-target optimization: one (training) is the stage with $|B|\gg|A|$, where $A$ includes the optimized targets and $B$ includes the unoptimized targets. The knowledge is transferred from the optimized one to nearby neighbors as a flooding algorithm, which is conducted efficiently through a D-level tree. The second stage (inference) is that $|A|\gg|B|$, each target $y_{k'}\in B$ can be also estimated from $A$, visualized as Figure~\ref{fig:overview} (a). The unoptimized (white dot) can learn from the nearest optimized target (black dot).

In this work, we will explain the theoretical foundations of MTQO, its structure in Section~\ref{sec:background}, and use techniques like the warm-start method and parameter estimator to enhance the MTQO process in Section~\ref{sec:proposed}. We also propose some deep learning techniques as more advantageous strategies. Finally, the initial comparative results are presented in Section~\ref{sec:experiment}. 


\section{Related works}

The popular technique in MTQO is the transfer optimization, which aims to improve learning performance, including the total number of optimization iterations and cost value, through the sharing of common knowledge between related targets. Similar principles from transfer optimization, such as evolutionary algorithms and Bayesian optimization, can be leveraged to enhance the performance of MTQO \cite{transfer_learning}. The knowledge transfer from a task $T_k$ can accelerate convergence on a target task $T_{k+1}$ in evolutionary algorithms ~\cite{7161358} or by transferring learned components of a surrogate model, such as priors and kernels in Bayesian optimization \cite{NIPS2013_f33ba15e}.

For the first issue, i.e., what and how to transfer, researchers have primarily focused on designing effective transfer mechanisms to share useful information among targets. Existing approaches can be referred to as multi-population-based methods. Multi-population-based methods evolve separate populations for each target and perform the transfer between populations based on evolutionary information, such as search distributions. Ref~\cite{8967000} proposed the multi-task genetic algorithm (MTGA) to minimize bias among target optima and to enable transfer through dimension-wise random shuffling. However, its random shuffling does not explicitly transfer optimization to account for dimensional similarity. Ref~\cite{LI20201555} proposed a multi-population framework with a novel mutation operator designed to enhance transferability between populations. More recently, research ~\cite{9627943} observed that transferring target-specific parameters alone may be insufficient and proposed a transfer approach that also transfers meta-knowledge between targets. This meta-knowledge captures higher-level strategies on how to obtain high-quality solutions and is thus more generalizable across targets with varying degrees of similarity.

Since target similarity is typically unknown a priori, transfer optimization may not always yield positive effects. An important research direction, therefore, involves estimating target similarity and encouraging transfer optimization when beneficial. For instance,~\cite{8666053} proposed a self-regulated EMTO algorithm that adjusts the intensity of transfer optimization based on observed target similarity during the search process. Ref~\cite{8672822} developed MFEA-II, incorporating an online parameter estimation strategy to assess target similarity and promote positive transfer optimization when targets are deemed similar.

\section{Quantum optimization algorithm}
\label{sec:background}

At present, most quantum optimization models are built upon parameterized quantum circuits (PQC)~\cite{benedetti2019parameterized, PhysRevA.106.052611}. PQCs serve as learnable models whose parameters can be trained via the parameter-shift rule (PSR), a general method for computing gradients in quantum systems \cite{PhysRevA.99.032331, Wierichs2022generalparameter}. Formally, the quantum optimization problem $P$ used in this research is defined as: given an ($L$ layers) trainable unitary $U(\bm\theta)$ where $\bm\theta^{(0)}$ is the initial parameter and a quantum state $|x\rangle$ with the cost function, for example:

\begin{align}\label{eq:cost_func}
    C(\bm\theta,x)= 1-\mathbf{Re}(|\langle\bm 0|U(\bm\theta)|x\rangle|^2).
\end{align}

We need to find the optimal parameter ${\bm\theta}^{*} \in \mathbb{R}^{d}$ such that:

\begin{align}\label{eq:argmin}
    \bm\theta^{*} = \text{argmin}_{\theta} C(\bm\theta,x).
\end{align}

The trainable quantum state is started with the $n$-qubit zero state $|\psi^{(0)}\rangle=|0\rangle^{\otimes n}$ and evolved at time step $t$ under $U(\bm\theta^{(t)})$:

\begin{align}
|\psi(\bm\theta^{t})\rangle=U(\bm\theta^{(t)})|\psi^{(0)}\rangle.
\end{align}

The target state $|x^* \rangle$ exists when the distance between it and the trainable state reaches $0$. Because $\mathfrak{Re}(|\langle\bm 0|U(\bm\theta)|x\rangle|^2)=1$ mean $|\psi^{(t)}\rangle$ and $|x\rangle$ are overlapped, thus $\mathcal{C}^*(x)\equiv\mathcal{C}(\bm\theta^*,x)\approx0$ is the minimal cost value. The Equation~\eqref{eq:argmin} is analogous to those commonly used in training classical machine learning models. The overall algorithm from~\eqref{eq:cost_func} to~\eqref{eq:argmin} is summarized in Algorithm.~\ref{algo:optimization}.

\begin{algorithm}[ht]
\caption{$P( \bm\theta^{(0)}, x)$} 
\label{algo:optimization}
\begin{algorithmic}[]
\Require $\bm\theta^{(0)},\bm x$
\For{$t$ in $[0,1,\ldots,n_{\text{iter}(x)}-1]$}
    \If{$\mathcal{C}(\bm\theta^{(t)}_{(x)}, x)>\tau$}
        \For{$j$ in $[0,1,\ldots,m-1]$}
            \State $\partial_{j}\mathcal{C}(\bm\theta^{(t)}_{(x)},x)\gets\text{PSR}(\mathcal{C},\bm\theta^{(t)}_{(x)},x,j)$
        \EndFor
        \State $\bm\theta^{(t+1)}_{(x)}\gets \text{Optimizer}(\nabla_{\bm\theta_{(x)}}\mathcal{C}(\bm\theta^{(t)}_{(x)},x))$ 
    \Else
        \State\Return $\bm\theta^{(t)}_{(x)}$
    \EndIf
\EndFor
\State\Return $\bm\theta^*\equiv\bm\theta^{(n_{\text{iter}})}$
\end{algorithmic}
\end{algorithm}

For each iteration $t$, the cost value $\mathcal{C}(\bm\theta^{(t)}_{(x)},x)$ is calculated from the quantum computer and then compared to a threshold $\tau\ll0$. If the PQC still has not converged, $\bm\theta^{(t)}_{(x)}$ needs to be updated through some first-order optimizers, such as Gradient Descent (GD) or Adam \cite{kingma2014adam}. This first-order optimizer requires the gradient defined as~\eqref{eq:gradient}:

\begin{align}\label{eq:gradient}
\nabla_{\bm\theta_{(x)}}&C(\bm\theta^{(t)}_{(x)},x)= \nonumber\\ &\left[\partial_{\theta_{0_{(x)}}}\mathcal{C}(\bm\theta^{(t)}_{(x)},x)\;\ldots\; \partial_{\theta_{m-1_{(x)}}}\mathcal{C}(\bm\theta^{(t)}_{(x)},x)\right]^{\intercal}.
\end{align}

The partial derivative $\partial_{\theta_{j_{(x)}}}\mathcal{C}(\bm\theta^{(t)}_{(x)},x)$ can be achieved via the PSR technique \cite{Wierichs2022generalparameter}. In case the generator of the quantum gate has two eigenvalues, such as one-qubit rotation gates, this technique is known as two-term PSR:

\begin{equation}
    \begin{aligned}
        \partial_{\theta_{j_{(x)}}}\mathcal{C}(\bm\theta^{(t)}_{(x)},x) &=
        -\partial_{\theta_{j_{(x)}}}  \mathcal{K}(\bm\theta^{(t)}_{(x)},x)\\
        &= 
     -\frac{1}{2}[\mathcal{C}(\bm\theta^{(t)}_{(x)} + \frac{\pi}{2} \bm e_j,x) \\ &\quad\; - \mathcal{C}(\bm\theta^{(t)}_{(x)} - \frac{\pi}{2} \bm e_j,x)],
\end{aligned}
\label{eq:psr2}
\end{equation}

$e_j$ is the $j^{\text{th}}$-unit vector. For each target $x$, Algorithm~\ref{algo:optimization} takes $n_{\text{iter}(x)}$ iterations;  each iteration refers to $2m+1$ quantum evaluations (QE) for cost value $\mathcal{C}(\bm\theta^{(t)}_{(x)},x)$ and its gradient. Totally, finding $\bm\theta^*_{(x)}$ require $n_{\text{iter}(x)}(2m+1)$ QEs. Note that $n_{\text{iter}(x)}$ grow exponentially by $n$.

\section{Strategies for multi-target quantum optimization}
\label{sec:proposed}

In many practical scenarios, optimization must be repeated across a family of input states where these inputs often reside in a continuous or structured space. Furthermore, solving~\eqref{eq:argmin} multiple times could lead to large resources on quantum computers.  Assuming that we have $K$ target in a same Hilbert space $\mathcal{S}$ that divided into two subsets: $A = \{ x_{(k)}\}_{k=1}^{K}$ and $B = \{y_{(k)}\}_{k'=1}^{K'}$, where $A$ includes the optimized targets and $B$ includes the unoptimized targets.
We transfer the knowledge from $A$ to $B$ such that $\sum_{k=1}^{K'}n_{\text{iter}(y_{k'})}$ reduce as much as possible. Specifically, given a new input $y_{k}\in B$, we define a distance metric $d(y_{k'}, x_k) $ on $\mathcal{S}$, and identify the nearest previously seen target $x_{k} $ such that:

\begin{align}\label{nearest_target}
    k^* = \text{argmin}_{k} d(y_{k'}, x_k).
\end{align}

The strategies based on this method can be arranged from simple to advanced. In most simple way, the target $k^*$ can be transferred to $x_{k'}$ if we use $\bm{\theta}^*_{(x_{k}^*)}$ as parameter initialization (warm-starting method). A bit more complicated, we can use an estimator to move the parameters closer to the optimal point. By using the above-defined metric, the unoptimized targets can also be organized for better optimization performance. Finally, we suggest that applying deep learning methods, which were successful in the classical multi-target optimization, may help us improve the performance of the quantum analogue.

\subsection{Warm-starting method}
\label{sec:warn-started}

If $P(\bm\theta^{(0)}, y_{k'})$ take $n_{\text{iter}(y_{k'})}$ iterations to converge. Since $ y_{k'} \approx x_{k^*} $ under the defined metric $d$, we expect $ \bm{\theta}^*_{(y_{k'})}$ to be close to $\bm{\theta}^*_{(x_{k^*})}$, which leads to a reduction in number of optimization iteration. On the other hand, $P(\bm\theta^*_{(x^*_{k})}, y_{k'})$ consume only  $n^*_{\text{iter}(y_{k'})}\ll n_{\text{iter}(y_{k'})}$ iterations. This approach can be viewed as a warm-starting optimization method \cite{Egger2021warmstartingquantum}. The procedure of this method is as follows. First, we optimize $\bm\theta_{(x)}$ for a target $x$ using randomly initialized parameters to minimize the cost value. Second, after the base PQC has been optimized, we fine-tune (continue to optimize) $\bm\theta^*_{(x)}$ to $\bm\theta^*_{(y)}$ for a new target $y$. Finally, we evaluate the effectiveness by comparing the performance of the PQC trained with the warm-starting and the PQC trained from scratch with default initialized parameters. 

\begin{figure*}[t]
\includegraphics[width=0.99\textwidth]{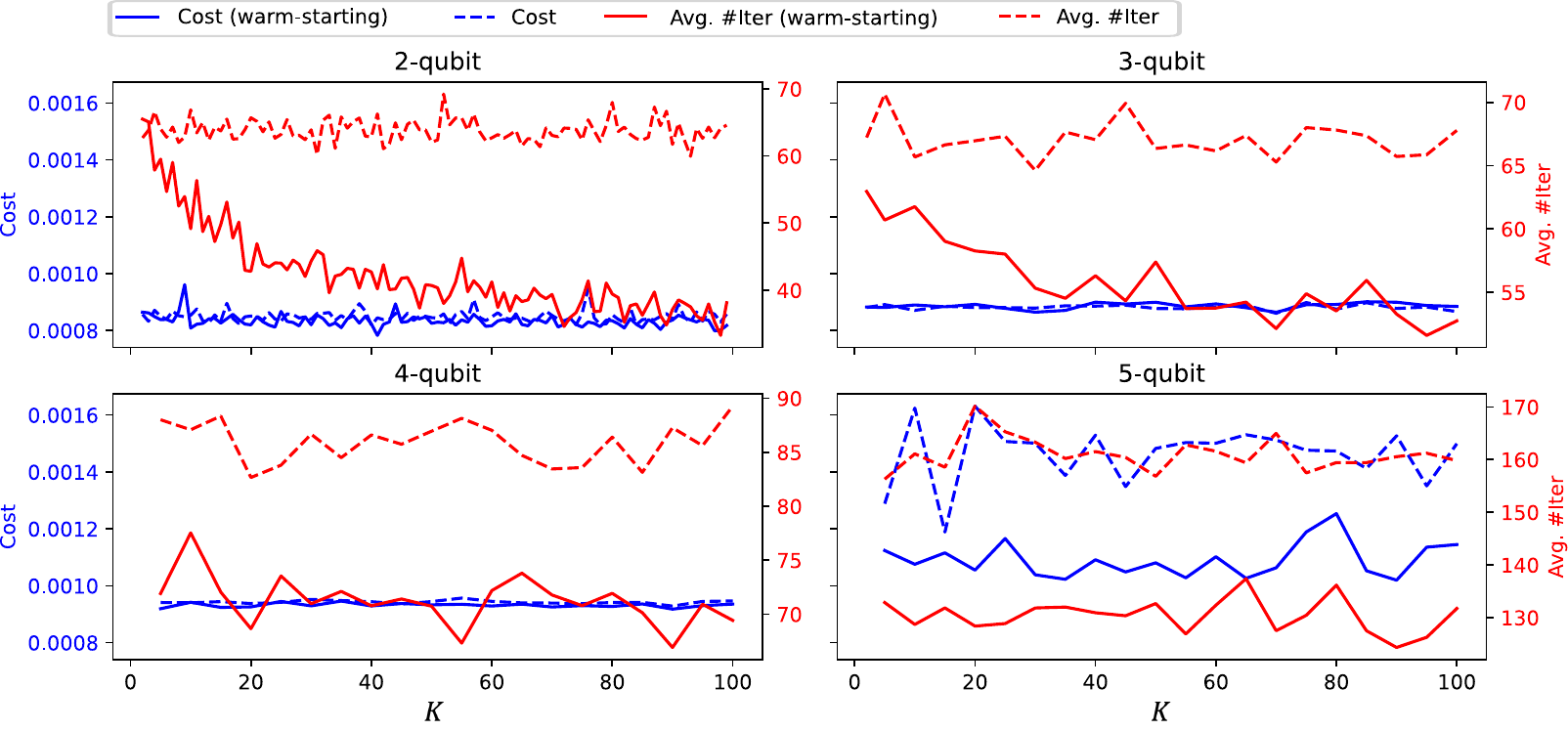}
\caption{The average of \#Iteration and cost $C_{(x)}(\bm\theta^*_{(x)})$. Each subplot represents one qubit configuration. The blue curves indicate the cost values obtained after optimization, while the red curves show the corresponding average of \#Iteration consumed to reach those cost values. Solid lines denote results with warm-starting, whereas dashed lines represent the baseline without warm-starting.}
\label{fig:cost_and_iter}
\end{figure*}

\subsection{Parameter estimator}

Rather than a warm-started meta-optimization method, we use a parameter estimator as introduced in~\cite{hai_image}:

\begin{align}\label{eq:estimator}
    \tilde{\mathfrak{E}}_{\bm\theta_{(x)} \to \bm\theta_{(y)}}(\bm\theta_{(x)}):\mathbb{R}^m\rightarrow\mathbb{R}^m
\end{align}

with the notation $\mathcal{C}_{(x)}(\bm\theta^{*}_{(x)})\equiv\mathcal{C}(\bm\theta^*_{(x)}, x)$ as the cost function for the target $x$. Let $\bm\theta^{*}_{(x)}$ be such that $\mathcal{C}_{(x)}(\bm\theta^{*}_{(x)})\approx0$. If the state is perturbed to $y = x + \epsilon$ with $\epsilon\approx0$, we consider the first-order Taylor expansion of $\mathcal{C}$ with respect to $\bm\theta$ at $\bm\theta^{*}_{(x)}$ and fixed $y$:

\begin{align}\label{eq:taylor_1}
\mathcal{C}_{(y)}(\tilde{\bm\theta}^{*}_{(y)}) &\approx \mathcal{C}_{(y)}(\bm\theta^{*}_{(x)}) \nonumber\\
&+ \nabla_{\bm\theta} \mathcal{C}_{(y)}(\bm\theta^{*}_{(x)})^\top \left( \tilde{\bm\theta}^{*}_{(y)} - \bm\theta^{*}_{(x)} \right).
\end{align}

Because $\tilde{\bm\theta}^{*}_{(y)}$ must satisfy $\mathcal{C}_{(y)}(\tilde{\bm\theta}^{*}_{(y)}) \approx 0$,~\eqref{eq:taylor_1} turn to: 
\begin{align}\label{eq:taylor2}
\mathcal{C}_{(y)}(\bm\theta^{*}_{(x)}) \approx - \nabla_{\bm\theta} \mathcal{C}_{(y)}(\bm\theta^{*}_{(x)})^\top \left( \tilde{\bm\theta}^{*}_{(y)} - \bm\theta^{*}_{(x)} \right).
\end{align}

Specifically, the pseudo-inverse of $\nabla_{\bm\theta} \mathcal{C}_{(y)}(\bm\theta^{*}_{(x)})$ is given by:

\begin{align}
\left[\nabla_{\bm\theta} \mathcal{C}_{(y)}(\bm\theta^{*}_{(x)})\right]^+ = \frac{\nabla_{\bm\theta} \mathcal{C}_{(y)}(\bm\theta^{*}_{(x)})^\top}{\|\nabla_{\bm\theta} \mathcal{C}_{(y)}(\bm\theta^{*}_{(x)})\|^2}.
\end{align}

Since this is a single scalar equation in the vector unknown $\tilde{\bm\theta}^{*}_{(y)}$, the corresponding parameters for $y$ are estimated as~\eqref{eq:approx}:

\begin{align}
\tilde{\bm\theta}^{*}_{(y)} \approx \bm\theta^{*}_{(x)} - \frac{\nabla_{\bm\theta} \mathcal{C}_{(y)}(\bm\theta^{*}_{(x)})}{\|\nabla_{\bm\theta} \mathcal{C}_{(y)}(\bm\theta^{*}_{(x)})\|^2}\, \mathcal{C}_{(y)}(\bm\theta^{*}_{(x)})
\label{eq:approx}
\end{align}

denoted as $\tilde{\bm\theta}^{*}_{(y)}=\tilde{\mathfrak{E}}_{\bm\theta_{(x)} \to \bm\theta_{(y)}}(\bm\theta^{*}_{(x)})$. If $\mathcal{C}_{(y)}(\tilde{\bm\theta}^{*}_{(y)})$ or $\mathcal{C}_{(y)}({\bm\theta}^{*}_{(x)})$ is less than $\tau$, $\tilde{\bm\theta}^{*}_{(y)}$or ${\bm\theta}^{*}_{(x)}$ is actually $\bm\theta^{*}_{(y)}$ , 
otherwise, $\bm\theta^{*}_{(y)}\gets P(\ldots)$ with $\bm\theta^{(0)}=\tilde{\bm\theta}^{*}_{(y)}$. Evaluating \eqref{eq:approx} take $2m+1$ quantum evaluation for $\nabla_{\bm\theta}\mathcal{C}_{(y)}(\bm\theta^*_{(x)})$ and $\mathcal{C}_{(y)}(\bm\theta^{*}_{(x)})$, respectively.

\subsection{Organizing targets as a multi-level tree}

A multi-level tree is a hierarchical approach to organizing data into clusters at multiple resolutions or abstraction levels. Formally, multi-level clustering constructs a sequence of $D$-partitions, as~\eqref{eq:cluster}:

\begin{align}\label{eq:cluster}
    {C}^{(1)}, {C}^{(2)}, \dots, {C}^{(D)}.
\end{align}

Each $\mathcal{C}^{(d)} = \{x^{(d)}, C^{(d-1)}_1, C^{(d-1)}_2, \dots, C^{(d-1)}_{|d|} \}$ include a centroid $x^{(d)}$ and the set of tree at lower level $d-1$. Conversely, trees at level $d$ are formed by merging clusters from level $d-1$. The lower levels capture finer-grained clusters and higher levels capture more abstract groupings. This structure allows analysis at multiple scales, facilitating both local and global interpretations of the data.

Figure~\ref{fig:overview} (b) presents the 2-level tree system where level 0 has two sub-trees. The knowledge can be transferred from the centroid to sub-trees, for example: knowledge from the $\bm\theta^*_{d=0}$ is transferred to $\bm\theta_{d=1}$, knowledge from the $\bm\theta^*_{d=1}$ is transferred to $\bm\theta_{d=2}$, and so on. The optimization process on tree can be imagined as Figure~\ref{fig:overview} (c), instead of starting from $\bm\theta^{(0)}$, $\bm\theta^{*}_{d=1}$ and $\bm\theta^{*}_{d=2}$ can be obtained from $\bm\theta^{*}_{d=0}$ and $\bm\theta^{*}_{d=1}$, respectively. For example, we could take a hundred iterations for achieving $\bm\theta^*_{d=1}$ but only a few iterations if we start from the nearest neighbor $\bm\theta^*_{d=0}$ instead of $\bm\theta^{(0)}$.

\subsection{Deep learning method}


The similarity between multi-target classical optimization and multi-target quantum optimization has spurred the integration of advanced machine learning paradigms to accelerate convergence and improve generalization across instances. Ref~\cite{11011037} study on transfer learning of PQC, the authors propose a mathematically grounded framework where a pretrained PQC from a source domain is adapted to a new target domain by analytically computing transitions in one‑parameter unitary subgroups, deriving an optimal parameter shift $\delta\theta^=(z^T z)^{-1} z^T q$ that minimizes expected fine‑tune loss in one shot rather than iterative gradient descent. This approach reduces the \#Iterations and circuit depth, effectively exhibiting a progressive learning pattern where knowledge from $A$ is reused to bootstrap performance on $B$. Similarly, hybrid models such as pre‑trained Tensor‑Train Network demonstrate pretraining of a classical TTN and subsequent embedding into a PQC, bounded by PL‑condition analysis to improve convergence and generalization—another form of progressive, step‑wise transfer learning~\cite{qi2023pre}.

In contrast, multi‑task learning and meta‑learning offer parallel and rapid adaptation strategies across multiple targets. According to classical multi‑task learning theory, joint optimization of several related tasks using a shared feature representation can induce useful inductive bias and improve generalization across tasks, especially under resource constraints or sparse sampling~\cite{strezoski2019taskrouting}. In quantum settings, this means if we train a PQC for task $k$, which is composed of many sub-trainable unitaries:

\begin{align}
    U(\bm\theta_k)=\bigotimes_{j=0}\left(U_{\{j\}}\left(\bm{\theta}_{k\{j\}}\right) \right).
\end{align}

$\bm\theta_{k\{j\}}$ is the parameter $k$ at layer $j$; we can share mid-layer parameter for $\bm\theta_{k+1}$. Meanwhile, meta‑learning methods enable fast adaptation: a meta‑learner network is trained across a distribution of $k''\ll K$ targets, producing initial PQC parameters that can be fine‑tuned in just a few gradient steps for each new target. Studies suggest that such meta‑strategies significantly improve robustness under noise \cite{Khairy_Shaydulin_Cincio_Alexeev_Balaprakash_2020, 10.5555/3305381.3305498}. Together, multi-task learning and meta‑learning comprise powerful tools for scalable, sample‑efficient optimization in quantum systems deploying multi‑target optimization.

\section{Experiments}
\label{sec:experiment}

We use $CR_{Y_\text{chain}}+ZXZ$ ansatz as $U(\bm\theta)$ in this research. This ansatz is separated as ($L$ times) entangled and rotation parts; these parts are made from $n$ control-rotation $Y$, $2n$ rotation $Z$, and $n$ rotation $X$.
The control-rotation gates are arranged following a chain (cycle) topology, formulated as~\eqref{eq:cry_chain}:

\begin{align}\label{eq:cry_chain}
U(\bm\theta) = ZXZ\prod_{i=0}^{n-1} CR_{Y}(\theta)_{i,\ (i+1)\%n}.
\end{align}

The number of parameters and circuit depth of the above $n$-qubit ansatz are $3nL$ and $6L$, respectively. The problem $P$ is implemented by Python 3.10 and Pennylane 0.41.0, and the scalability is investigated by increasing $n$ and $L$ ($L=n$). The maximum \#Iteration and $\tau$ are fixed at 200 and $10^{-3}$ for all cases, respectively. Higher number of qubits, harder to optimize; in other words, $P$ will require more iterations to achieve convergence. The dataset is generated quite simply; we randomly choose $k$ $n$-qubit quantum states, then sequentially optimizing each of them with $P(\bm\theta^{(0)}, x)$ as defined in Algorithm~\ref {algo:optimization}. The final cost values $\mathcal{C}$ conducted by our proposed methods and baseline method are compared in the following section.

\subsection{Effectiveness of transfer-based strategies}

\begin{table}[t]
\centering
\caption{Comparison of warm-starting method and parameter estimator versus the no acceleration technique.}
\label{tab:summary}
\resizebox{0.48\textwidth}{!}{%
\begin{tabular}{|c|l|l|l|l|}
\hline
\multicolumn{5}{|c|}{\textbf{Warm-starting method}} \\
\hline
$n$ & Cost (Proposed) & Cost & Avg.\#Iter (Proposed) & Avg.\#Iter \\
\hline
2 & 0.0008 & 0.0008 & 42.41$\;\downarrow$ & 63.80 \\
3 & 0.0009 & 0.0009 & 55.92$\;\downarrow$ & 67.05 \\
4 & 0.0009 & 0.0009 & 71.25$\;\downarrow$ & 85.95 \\
5 & 0.0011$\;\downarrow$ & 0.0015 & 130.44$\;\downarrow$ & 161.01 \\
6 & 0.0153$\;\downarrow$ & 0.0188 & 199.00 & 199.00 \\
7 & 0.0874$\;\downarrow$ & 0.0940 & 199.00 & 199.00 \\
\hline
\multicolumn{5}{|c|}{\textbf{Parameter estimator method}} \\
\hline
$n$ & Cost (Proposed) & Cost & Avg.\#Iter (Proposed) & Avg.\#Iter \\
\hline
2 & 0.0008$\;\downarrow$ & 0.0009 & 34.48$\;\downarrow$ & 65.63 \\
3 & 0.0009 & 0.0009 & 49.05$\;\downarrow$ & 67.25 \\
4 & 0.0009 & 0.0009 & 69.57$\;\downarrow$ & 85.96 \\
5 & 0.0011$\;\downarrow$ & 0.0012 & 135.84$\;\downarrow$ & 160.50 \\
6 & 0.0155$\;\downarrow$ & 0.0185 & 197.04$\;\downarrow$ & 197.31 \\
7 & 0.0888$\;\downarrow$ & 0.0939 & 196.98$\;\downarrow$ & 197.45 \\
\hline
\end{tabular}%
}
\end{table}

Table~\ref{tab:summary} summarizes all comparison results between transfer-based strategies and non-transfer-based strategies approaches, for two distinct methods: the warm-starting and the parameter estimator method. For each setting, we report the average of minimum cost values and the average number of iterations (Avg.\#Iter) required to reach convergence. The warm-starting method is benchmarked as follows: given sets $A$ and $B$ where $|A|\gg|B|$, for each $y\in B$, we optimize $y$ with and without using knowledge from $A$. On the other hand, for the parameter estimator, we start with $B$ and an empty set $A$; for each optimized target, we move it from $B$ to $A$ until $B$ becomes an empty set. The first target will be optimized without the parameter estimator, and the latter will be transferred from $5$ nearest targets.

In both warm-starting and parameter estimator methods, transfer-based strategies consistently reduce \#Iterations required, particularly in lower qubit regimes. For instance, at 2-qubit, the warm-starting method reduces the average \#Iterations from $63.80$ to $42.41$, representing a significant acceleration in convergence without compromising final cost (both at $0.0008$). Similar behavior is observed up to 5-qubit, where the warm-starting method consistently achieves comparable or better cost values with fewer iterations. As the number of qubits increases $5$, the iteration counts approach the maximum limit (200 iterations), suggesting that the optimization reaches a computational cap. Nevertheless, both method still yields slightly better cost values, reducing the cost in the warm-starting method from $0.0188$ to $0.0153$ at 6-qubit, and from $0.0940$ to $0.0874$ at 7-qubit.

\begin{figure}[t]
\includegraphics[width=0.49\textwidth]{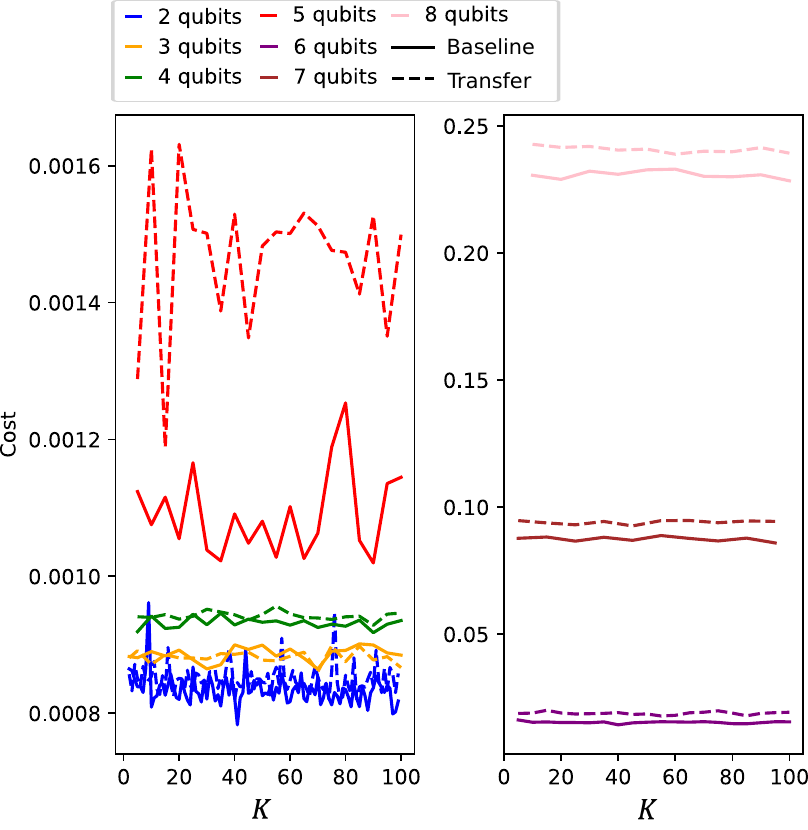}
\caption{The cost value between no warm-starting (normal line) and warm-starting (dotted line), from 2-qubit to 8-qubit cases. }
\label{fig:cost_only}
\end{figure}

Figure~\ref{fig:cost_and_iter} shows the detailed result from the warm-starting method from 2-qubit to 5-qubit. Across all $K$, our method consistently achieves lower or comparable cost values using fewer iterations. In the 2-qubit case, the cost with warm-starting converges rapidly to a stable lower value, while the required \#Iterations is significantly reduced compared to the baseline. This benefit is particularly pronounced for smaller systems, where the prior knowledge transferred from related tasks yields substantial optimization efficiency. In the 3-qubit and 4-qubit cases, the \#Iteration for the warm-starting is again consistently lower, while both approaches maintain similar cost levels. For 5-qubit, as system complexity increases, the benefits of warm-starting become clearer, where both cost and \#Iteration are better. However, even in these cases, warm-starting demonstrates a trend toward requiring fewer iterations to reach comparable or slightly better performance, indicating its scalability. In higher-dimensional systems (6 to 8 qubits, Figure~\ref{fig:cost_only}, right panel), the advantage of warm-starting becomes more pronounced. The dashed lines lie consistently below the solid lines, indicating lower final cost values even in more complex quantum systems. Furthermore, the cost values obtained via transfer are relatively stable as $K$ increases, in contrast to the baseline, which exhibits higher variability. This highlights the robustness of warm-starting in high-dimensional optimization landscapes.

\subsection{Limitation based on the number of qubits}

\begin{figure}[t]
\includegraphics[width=0.49\textwidth]{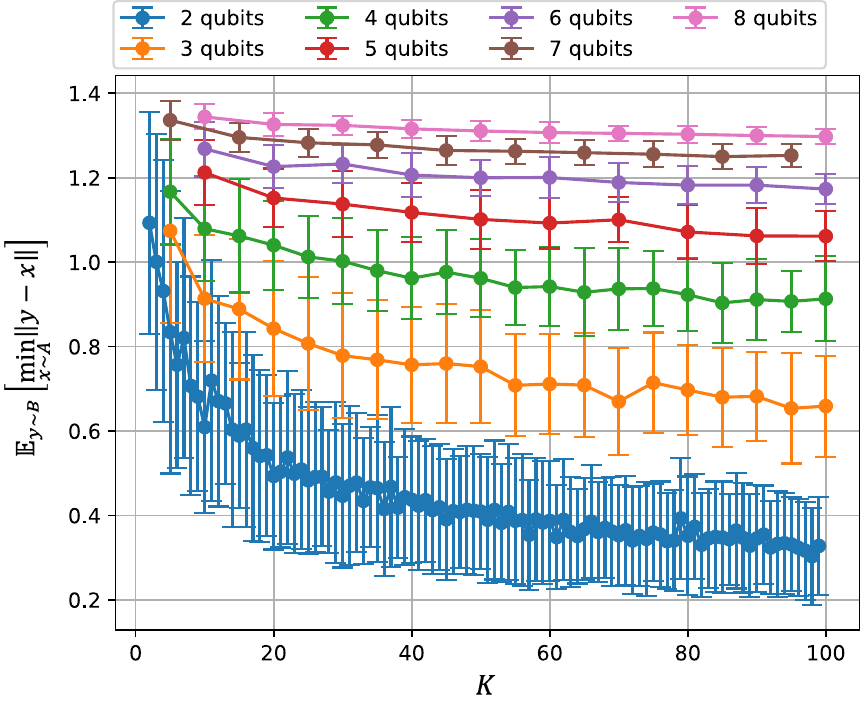}
\caption{Expected minimum distance between a target $y\in B$ and $A$, we investigate from 2-qubit to 8-qubit. The error bar is plotted with the mean and standard deviation from 100 trials.}
\label{fig:distance}
\end{figure}

The average distance from set $B$ to set $A$ according to the shortest distance (nearest neighbor) is denoted by the expectation operator as follows:

\begin{align}
d_{\text{avg-min}}(B, A) &= \mathbb{E}_{y \sim B} \left[ \min_{x \in A} \|y-x\| \right] \nonumber\\
&= \frac{1}{K'} \sum_{k'=1}^{K'} \min_{k = 1, \dots, K} \| y_{k' } - x_k \|
\end{align}

where the distance between any two target $\{x,y\}$ is simply defined as the norm-2, $\|y-x\|\equiv \|y-x\|_2=2-2\mathbf{Re}(\langle x|y\rangle)$. The distance metric is norm-2 instead of fidelity in case we want to control the density of the generated dataset. Figure~\ref{fig:distance} illustrates the expected minimum distance between a new target $y \sim \mathcal{B}$ and a set of $K$ previously targets $x \in \mathcal{S}$. The vertical axis shows the quantity $\mathbb{E}_{y \sim B} \left[ \min_{x \in A} \|y-x\| \right]$, measuring how close a new target is, on average, to its nearest neighbor in $A$. The horizontal axis corresponds to $K=|A|$. Across all cases from 2 to 8-qubit, increasing $K$ leads to a monotonic decrease in the expected minimum distance, as a larger reference set increases the probability of locating a closer neighbor. However, this decay saturates as $K$ grows, indicating diminishing returns with additional targets.

Furthermore, the distances increase with the number of qubits. For example, the 2-qubit case achieves significantly lower average distances than the 8-qubit case at all values of $k$. This trend is consistent with the curse of dimensionality: as the dimension of $\mathcal{S}$ increases, random vectors become increasingly distant, leading to sparser convergence of the space even at higher $K$. The error bars are largest at low $K$, especially for lower-qubit configurations, reflecting high variability when the reference set is sparse. These results underscore the growing challenge of generalization and interpolation in high-dimensional quantum state spaces. In particular, nearest-neighbor-based retrieval becomes less effective with increasing qubit number, motivating the need for more expressive models. So, the warm-starting method becomes unsuitable for large-qubit optimization since $K$ must increase exponentially with the dimension of $S$.

\section{Conclusion}
In this work, we formally introduced the problem of multi-target quantum optimization (MTQO) and proposed a general framework that integrates some initial strategies for better performance. By organizing the optimization space into structured sets and applying knowledge transfer techniques, such as warm-starting and parameter estimators, we demonstrated that optimization for new targets can be significantly accelerated. Experimental evaluations show that by applying different techniques, we can consistently reduce the number of iterations required and achieve comparable or better final cost values across a range of qubit configurations.

Our results suggest that the effectiveness of transfer increases with the number of pre-optimized targets and that using appropriate similarity metrics is critical for effective knowledge transfer. However, the diminishing returns observed in higher-dimensional cases, due to the curse of dimensionality, indicate limitations when scaling to larger systems. Consequently, while warm-start and neighbor-based estimators are practical for small- and mid-sized circuits, alternative strategies such as meta-learning or hierarchical learning are needed for high-qubit regimes.

This work provides a foundation for scalable MTQO on near-term quantum devices and opens several avenues for future research, including optimal metric design, adaptive clustering, and integration with classical surrogate models. As quantum hardware continues to evolve, such transfer-based optimization frameworks could play a key role in maximizing efficiency and performance across complex quantum tasks. Although MTQO can be viewed as an analogue to classical multi-target optimization, it presents unique challenges. Several open questions remain for future exploration, including: (1) Because the distance decides how we organize and determine the best optimized target for transfer, which metric should we use? (2) How does the structure of $U$ affect the performance of transfer techniques?
(3) Will the deep learning method improve the performance on PQCs? and (4) How are the multi-target classical optimization and MTQO different?

\section*{Code availability}

The codes used for this study are available at https://github.com/NAIST-Archlab/MTQO. 

\section*{Acknowledgment}

This work was supported in part by the Japan Science and Technology Agency (JST)-Advanced Technologies for CArbon-Neutral (ALCA-Next)-Next Program, Japan, under Grant JPMJAN23F4; in part by the Japan Society for the Promotion of Science (JSPS), Grants-in-Aid for Scientific Research (KAKENHI), Japan, under Grant 22H00515; in part by the 2025 Priority Strategic Funds Promotion of open access publication of scholarly articles; and in part by the Next Generation Researchers Challenging Research Program under Grant zk24010019.

\bibliographystyle{IEEEtran}
\bibliography{references.bib}

\end{document}